\documentclass[aps,pra,
epsfigure,twocolumn,longbibliography,superscriptaddress]{revtex4-1}

\usepackage{amsfonts}
\usepackage{amssymb}
\usepackage{amsmath}
\usepackage{calc}
\usepackage{subfigure}
\usepackage{graphicx}
\usepackage{epstopdf}
\usepackage{dcolumn}
\usepackage{bm}
\usepackage{color} 
\usepackage{txfonts}
\usepackage[dvipsnames]{xcolor}
\usepackage{appendix}
\usepackage{float}
\usepackage[colorlinks, citecolor=blue]{hyperref} 

\newcommand{\beq}{\begin{equation}}
\newcommand{\eeq}{\end{equation}}
\newcommand{\beqa}{\begin{eqnarray}}
\newcommand{\eeqa}{\end{eqnarray}}

\begin{document}

\title{Smooth bang-bang shortcuts to adiabaticity for atomic transport in a moving harmonic trap}

\author{Yongcheng Ding}
\email{jonzen.ding@gmail.com}
\affiliation{International Center of Quantum Artificial Intelligence for Science and Technology (QuArtist) \\ and Department of Physics, Shanghai University, 200444 Shanghai, China}
\affiliation{Department of Physical Chemistry, University of the Basque Country UPV/EHU, Apartado 644, 48080 Bilbao, Spain}

\author{Tang-You Huang}
\affiliation{International Center of Quantum Artificial Intelligence for Science and Technology (QuArtist) \\ and Department of Physics, Shanghai University, 200444 Shanghai, China}

\author{Koushik Paul}
\affiliation{International Center of Quantum Artificial Intelligence for Science and Technology (QuArtist) \\ and Department of Physics, Shanghai University, 200444 Shanghai, China}

\author{Minjia Hao}
\affiliation{International Center of Quantum Artificial Intelligence for Science and Technology (QuArtist) \\ and Department of Physics, Shanghai University, 200444 Shanghai, China}

\author{Xi Chen}
\email{xchen@shu.edu.cn}
\affiliation{International Center of Quantum Artificial Intelligence for Science and Technology (QuArtist) \\ and Department of Physics, Shanghai University, 200444 Shanghai, China}
\affiliation{Department of Physical Chemistry, University of the Basque Country UPV/EHU, Apartado 644, 48080 Bilbao, Spain}

\date{\today}

\begin{abstract}
	
Bang-bang control is often used to implement a minimal-time shortcut to adiabaticity for efficient transport of atoms in a moving harmonic trap. However, drastic changes of the on-off controller, leading to high transport-mode excitation and energy consumption, become infeasible under realistic experimental conditions. To circumvent these problems, we propose smooth bang-bang protocols with near-minimal time, by setting the physical constraints on the relative displacement, speed, and acceleration between the mass center of the atom and the trap center. We adopt Pontryagin's maximum principle to obtain the analytical solutions of smooth bang-bang protocol for near-time-minimal control. More importantly, it is found that the energy excitation and sloshing amplitude are significantly reduced at the expense of operation time. We also present a multiple shooting method for the self-consistent numerical analysis. Finally, this method is applied to other tasks, e.g., energy minimization, where obtaining smooth analytical form is complicated. 

\end{abstract}

\maketitle

\section{introduction}

\label{sec:intro} 

Precise control and manipulation of ultracold atomic systems without excitation or loss are challenging and important for the practical applications in atom interferometry, quantum-limited metrology, and quantum information processing ~\cite{benkish2002ion,hansel2001bose,hansel2001prl,gustavson2001prl,reichle,spaceborneBEC,schwartz,klitzingnature2019}. For example, protocols with existing adiabatic methods have been well developed  to transport cold atoms by various traps ~\cite{denschlagnjp2006,davidpra2006,wangapl2010,jamespra2011,winelandprl2012,poschingerprl2012,sterrnjp2012,homenjp2013,gaaloulnjp2018}. However, the operation time required for approximating adiabatic processes is much longer than decoherence time, which may ruin the desired results in practice. To remedy it, several approaches, including but not limited to Fourier method \cite{davidepl2008,davidandmuga}, optimal control theory~\cite{calarcopra2009,calarocqip2013,hessmosp2017} and machine learning~\cite{shersonnature2016,sels}, have been attempted to reduce the timescales beyond the adiabatic limits.

Over the last decade, the concept of ``shortcuts to adiabaticity" (STA)~\cite{chenprl2010} provides an alternative approach for speeding up adiabatic processes without residual energy excitation in various quantum systems (see review articles~\cite{review1,review2}). The most common methods include fast-forwarding scaling~\cite{masuda2010,masudapra2012}, counter-diabatic driving~\cite{prx,kimnc2016}, and invariant-based inverse engineering~\cite{erikpra2011,eriknjpbec,mikelpra2013,mikelpra2014,tobalina}, in which the adiabatic transport is accelerated by modifying the trap trajectory or introducing auxiliary interaction to compensate the inertial force. Among them, inverse engineering, combined with perturbation theory or/and optimal control, is capable of designing the optimal shortcuts with transport time~\cite{chenpra2011,kochnjp}, energy excitation \cite{charronsp2019}, anharmonic effect~\cite{qipra2015,qijpb2016,jing}, fluctuating trap frequency and position~\cite{xiaojing14,xiaojing15,xiaojing18}. As expected from Pontryagin's maximum principle, bang-bang control is indeed the time-optimal solution of atomic transport with harmonic traps ~\cite{chenpra2011,kochnjp}. However, it has been observed that the abrupt change of the control function at the switching points has severe consequences for practical implications. For instance, as seen in Ref.~\cite{nessnjp2018}, this leads to the excitation of dynamical modes around the switching points which violates the fundamental assumption of STA methods of constraining the system in a particular mode during the evolution. In addition, the onset of step function entails sudden control over position, velocity, and acceleration of the trap by a spatial light modulator, which makes the experiment complicated. Therefore an effective and continuous control of the physical constraints of trap velocity ~\cite{stefanatosieee} or acceleration is essential that suppresses the energy excitation by protracting the process as a trade-off. 

In this paper, we present a study on near-minimal-time transport of cold atoms with a moving trap by combining inverse engineering and optimal control theory. Previous research~\cite{chenpra2011} suggests that the bounded controller for time-optimal transport should be of bang-bang type, which maximizes the control Hamiltonian following Pontryagin's maximum principle.
Here we focus on the smooth bang-bang trajectories by setting up more constraints that bound the first- and second-order derivatives of the control input, describing the relative velocity and acceleration. We verify that the energy excitation and sloshing amplitude can be significantly reduced by smooth bang-bang protocols, while the minimal timescale is slightly increased. Since the analytical expressions of trap trajectories become more complicated when the higher-order derivatives of the controller are bounded, we introduce a multiple shooting method~\cite{bassam}, as a numerical approach, to confirm the analytical results. Additionally, this numerical method can be further exploited to minimize other target functionals, e.g., time-averaged potential energy, where finding an analytical solution might pose difficulties. Finally, we emphasize that our results can be extended to other scenarios~\cite{stefanatospra2010,xiaojingpra2014,kosloff,freericksbangbang,freericksbangbang2} without loss of generality.

\section{Hamiltonian and Model}
\label{sec:model}

\begin{figure}[]
	\includegraphics[width=8.0cm]{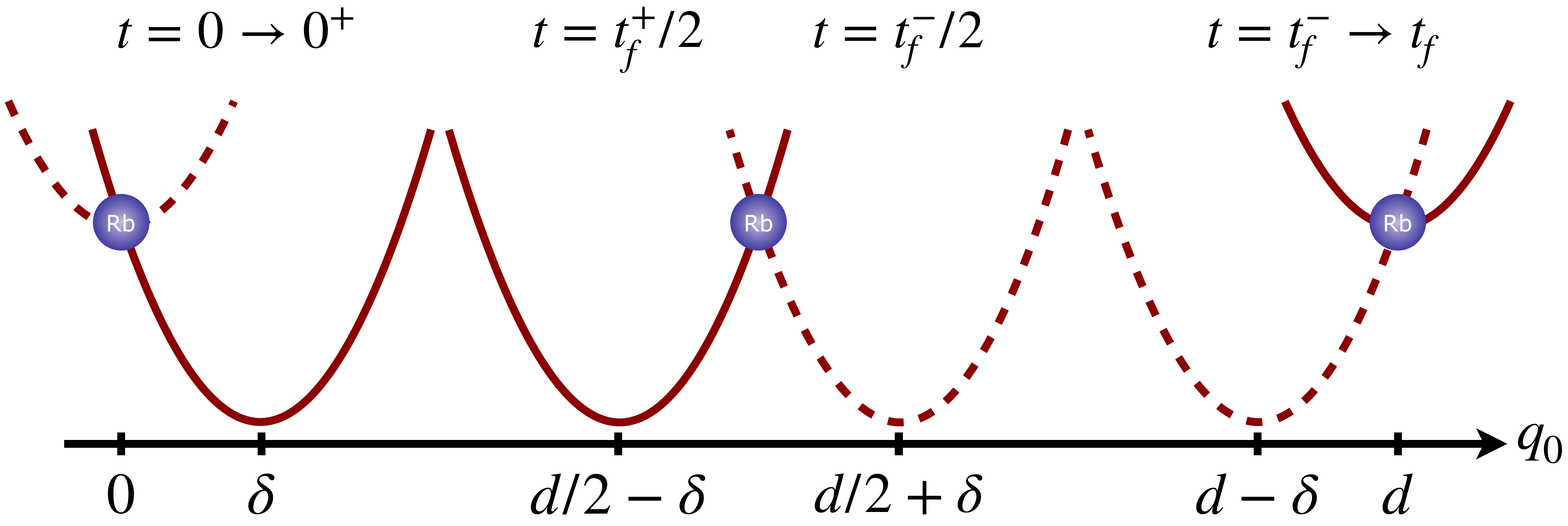}
	\caption{Schematic diagram of STA transport of a $^{87}$Rb atom in a moving harmonic trap, in which the relative displacement between trap center and mass center of the cold atom is changed from negative and positive values, for instance, in bang-bang control. The controller is bounded by $\delta$, transporting the atom for a distance of $d$ within minimal time $t_f$.}
	\label{Fig.scheme}
\end{figure}

For simplicity, we consider the time-dependent Hamiltonian that describes the transport of a single atom trapped in a rigid harmonic trap (see Fig.~\ref{Fig.scheme}),
with center $q_0(t) \equiv q_0$ and trap frequency $\omega_0$, which reads as
\begin{equation}
\label{Ht}
H(t)=\frac{\hat{p}^2}{2m}+\frac{1}{2}m\omega_0^2[\hat{q}-q_0(t)]^2,
\end{equation}
where $\hat{p}$ and $\hat{q}$ are momentum and position operators.
Eq. (\ref{Ht}) provides a good approximation for optical dipole interaction in low temperatures as one could easily neglect the effect of the anharmonic terms \cite{qijpb2016,nessnjp2018}.
This single-particle Hamiltonian 
possesses a quadratic-in-momentum Lewis-Riesenfield  invariant \cite{erikpra2011,LRO,LR1,LR2},
\begin{equation}
\label{It}
I(t)=\frac{1}{2m}(\hat{p}-m\dot{q}_c)^2+\frac{1}{2}m\omega_0^2[\hat{q}-q_c(t)]^2,
\end{equation}
where the parameter $q_c(t) \equiv q_c$ satisfies the auxiliary equation
\begin{equation}
\label{aeq}
\ddot{q}_c+\omega_0^2(q_c-q_0)=0,
\end{equation}
to guarantee self-consistency because of the invariant condition
\begin{equation}
\label{ic}
\frac{dI(t)}{dt}\equiv\frac{\partial I(t)}{\partial t}+\frac{1}{i\hbar}[I(t),H(t)]=0.
\end{equation}
Coincidentally, Eq.~\eqref{aeq} has the same structure of Newton's equation that governs the dynamics of a classical harmonic oscillator. Transport modes are described as 
\begin{eqnarray}
\psi_n(q,t)=e^{i\frac{m\dot{q}_cq}{\hbar}} \phi_n(q-q_c),
\label{ev}
\end{eqnarray}
where $\phi_n$ are the eigenstates of a static harmonic oscillator. The solution of the time-dependent Schr\"odinger equation, $i \hbar \partial_t \Psi(q,t)= H(t) \Psi(q,t)$, is constructed as the superposition of transport modes,
$\Psi(q,t)=\sum_n c_n \exp(i \alpha_n)\psi_n (q,t)$,
where $c_n$ are the time-independent coefficients, and
$\psi_n(q,t)$ are the eigenstates of dynamical invariant $I(t)$. Here the eigenvalues $\lambda_n$, satisfying $I(t) \psi_n (t) = \lambda_n \psi_n(t)$, are constants and the Lewis-Riesenfield phase $\alpha_n$ is calculated as
\begin{equation}
\alpha_n(t)=-\frac{1}{\hbar}\int_0^t [(n+\frac{1}{2})\hbar\omega_0+\frac{1}{2}m\dot{q}_c^2] dt',\label{LRphaseS}
\end{equation}
It is noted that all transport modes are orthogonal to each other at any time, being centered at $q_c(t)$. 

For a transport mode, the instantaneous average energy, $E(t)= \langle \Psi(t)|H(t)|\Psi(t)\rangle$, is calculated as~\cite{chenpra2011},
\begin{equation}
E(t)=\hbar\omega_0(n+\frac{1}{2})+\frac{m}{2} \dot{q}_c^2+ \frac{m}{2} \omega_0^2(q_c-q_0)^2,
\label{iaE}
\end{equation}
where the instantaneous average potential energy reads, 
\begin{equation}
V(t)=\hbar\omega_0(n+\frac{1}{2})+ E_p.
\label{iaE}
\end{equation}
The first term refers to constant ``internal" contribution. The second term $E_p=  m \omega_0^2(q_c-q_0)^2/2$ shares the form of a potential energy for a classical particle. Intuitively, high potential energy results in easy escape of a cold atom from the anharmonic trap in practice, reducing the effectiveness of STA
 \cite{qijpb2016}. In order to characterize the energy excitation for the whole process, we finally write down the time-averaged potential energy,
\begin{equation}
\label{energy1}
\bar{E}_p \equiv \frac{1}{t_f} \int_{0}^{t_f} E_p dt = \frac{1}{t_f}\int_{0}^{t_f} \frac{m}{2} \omega_0^2(q_c-q_0)^2 dt,
\end{equation}
as a consequence.

In addition, we are also interested in sloshing amplitude $\mathcal{A}$,
\begin{equation}
\label{energy2}
\mathcal{A} (t_f)= \left|\int_{0}^{t_f} \dot{q}_0(t) e^{- i \omega_0 t'} dt'\right|,
\end{equation}
which is the Fourier component at the trap frequency of the trap velocity trajectory. Nullifying the sloshing amplitude provides the optimal trajectory in the anharmonic case, thus improving the performance of STA in a realistic experiment \cite{nessnjp2018}.

In order to design the optimal trajectory of the harmonic trap by inverse engineering as usual, we suppose that the harmonic trap moves from $q_0(0)=0$ to $q_0(t_f)=d$ at finite shortened time $t_f$. To avoid final energy excitation, boundary conditions
\begin{eqnarray}
q_c(0)=0;~ \dot{q}_c(0)=0;~  \ddot{q}_c(0)=0,
\label{bc1}
\\
q_c(t_f)=d;~  \dot{q}_c(t_f)=0;~  \ddot{q}_c(t_f)=0,
\label{bc2}
\end{eqnarray}
are imposed along with Eq.~\eqref{aeq}. In addition,  the boundary conditions 
\begin{eqnarray}
\dddot{q}_c(0)=0;~ ~  \dddot{q}_c(t_f)=0,
\label{bc3}
\end{eqnarray}
are introduced to eliminate sloshing amplitude $\mathcal{A} (t_f)$ for encapsulating the energy in transport modes. Here we give an example of a simple polynomial \textit{Ans\"atz}, interpolating the center of transport modes, 
\beq
\label{poly}
q_c(t) = d\left[35 s^4-84s^5+70s^6-20s^7\right],
\eeq
originally proposed by Ref.~\cite{nessnjp2018},
with $s=t/t_f$. Once $q_c(t)$ and transport time $t_f$ are fixed, the optimal trajectory of the harmonic trap can be given by Eq.~\eqref{aeq}. However, we notice that this \textit{Ans\"atz} is not optimized enough, which will be analyzed by numerical results below. 

\section{Smooth bang-Bang Control with Near-minimal-time}

In this section, we use Pontryagin's maximum principle~\cite{book} for solving the near-minimal-time transport problem, leading to smooth bang-bang control. In general, the time-dependent control function $u(t)$ for minimizing the cost functional,
\beq
J(u)=\int_0^{t_f}g[\textbf{x}(t),u]dt,
\eeq
can be solved by constructing the following control Hamiltonian
\beq
H_c[\textbf{p}(t),\textbf{x}(t),u]=p_0g[\textbf{x}(t),u]+\textbf{p}^T\cdot\textbf{f}[\textbf{x}(t),u],
\eeq
where for the dynamical system $\dot{\textbf{x}}=\textbf{f}[\textbf{x}(t),u]$, the extremal solutions satisfy the canonical equations
\beqa
\dot{\textbf{x}}=\frac{\partial H_c}{\partial \textbf{p}},~~~
\dot{\textbf{p}}=-\frac{\partial H_c}{\partial \textbf{x}}.
\label{canonicaleqp}
\eeqa
Here the corresponding adjoint state $\textbf{p}$ formed by Lagrange multipliers, where $p_0 < 0$ can be chosen for convenience, and all components are nonzero and continuous, is such that $H_c$ reaches its maximum at $u\equiv u(t)$ for almost all $0\leq t \leq t_f$. More specifically, 
to find the time-optimal problem, we define the cost functional
\begin{equation}
J_T =\int_0^{t_f} 1 dt,
\label{J}
\end{equation}
and a control Hamiltonian,
\begin{equation}
H_c[\textbf{p}(t),\textbf{x}(t),u]=p_0+\textbf{p}^T\cdot\textbf{f}[\textbf{x}(t),u],
\label{Hc}
\end{equation}
where the dynamical system $\dot{\textbf{x}}=\textbf{f}[\textbf{x}(t),u]$ is governed by Eq.~\eqref{aeq}.

\subsection{bang-bang time-optimal control}

\begin{figure}[]
	\includegraphics[width=8.0cm]{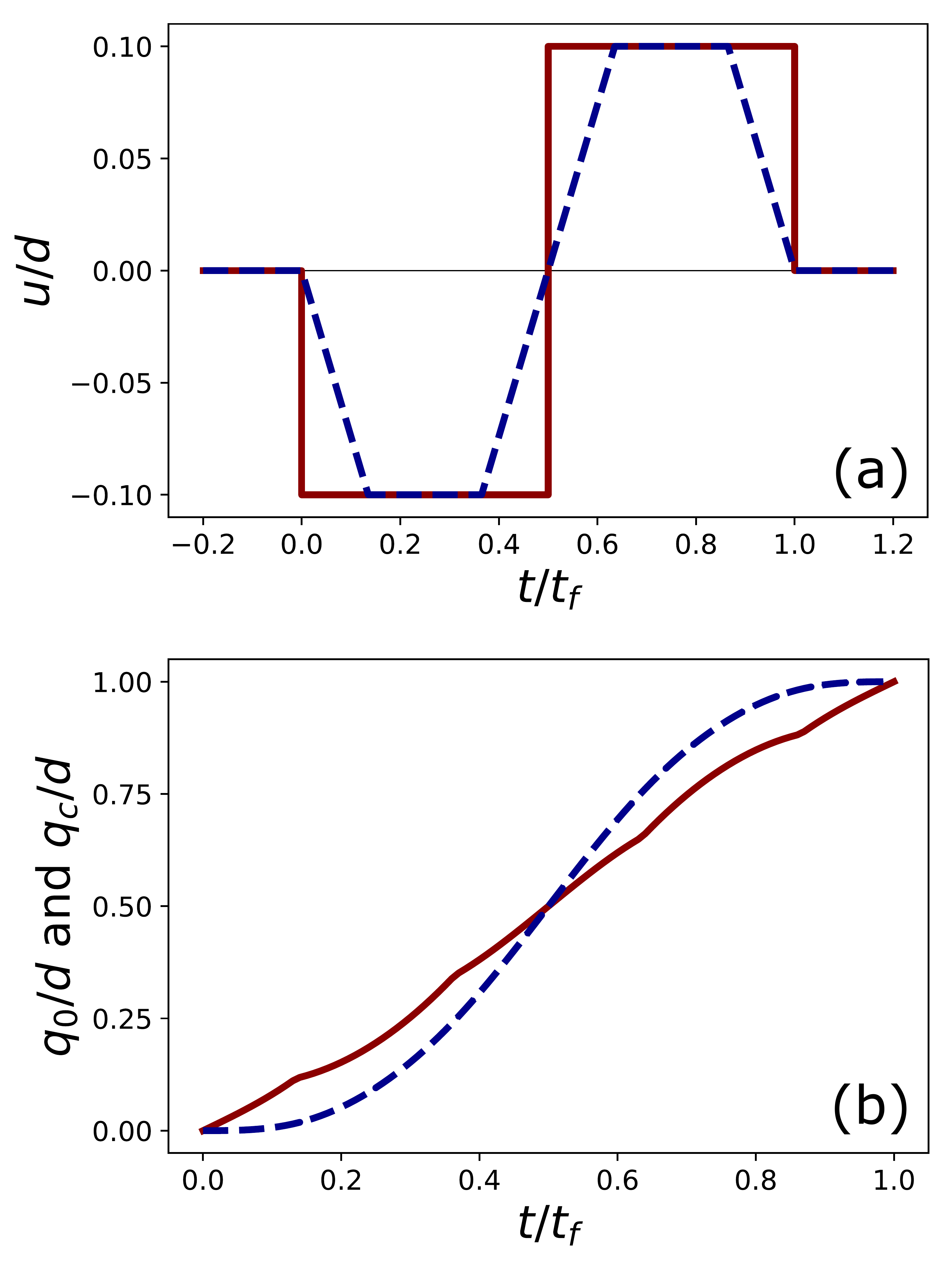}
	\caption{(a) Bang-bang-type controller $u(t)$ (solid red) and  smooth bang-bang controller $u(t)$ (dashed blue) with a constrained relative displacement $\delta$ and velocity $\epsilon$. (b) Smooth trajectories of the trap center $q_0(t)$ (solid red) and the mass center $q_c(t)$ of a cold atom (dashed blue). Parameters: trap frequency $\omega_0=2\pi\times20$ Hz; distance of transporting $d=0.01$ m; $m=1.44269 \times 10^{-25}$ kg, the mass of  $^{87} \mbox{Rb}$ atoms; the constraint on the relative displacement $\delta/d=0.1$;  and velocity $\epsilon/(d \omega_0)=0.1$. The corresponding near-minimal time \eqref{eqtfv} is $t_f=58.9$ ms, slightly larger than the minimal time $t_f^{\min}= 50.3$ ms, given by Eq.~\eqref{mint}.}
	\label{fig2}
\end{figure}

Let us first review the time-optimal control with bounded relative displacement to establish the background for analyzing smooth bang-bang control. By introducing a new notation,
\begin{equation}
x_1=q_c, x_2=\dot{q}_c, u(t)=q_c-q_0,
\label{newnotation-bangbang}
\end{equation}
we reformulate Eq.~\eqref{aeq} in the language of optimal control theory as follows,
\begin{eqnarray}
\dot{x}_1&=&x_2,
\\
\dot{x}_2&=&-\omega_0^2 u(t),
\label{aeqnn-bangbang}
\end{eqnarray}
where $x_{1,2}$ are the components of state vector $\textbf{x}$, and $u(t)$ is the scalar control function. Due to the anharmonicity of traps \cite{qijpb2016}, relative displacement $u(t)$ should be bounded by $|u(t)|\leq \delta$. Hence, the time optimization problem essentially comes down to the cost functional $J_T$ [see Eq.~\eqref{J}],
under the constraint $|u(t)|\leq\delta$. With the boundary conditions, $u(0)=u(t_f)=0$, the transport process occurs between $x_1(0)=0$ and $x_1(t_f)=d$ while $x_2(0)=x_2(t_f)=0$. The control Hamiltonian  (\ref{Hc}) for such choices can be written as
\begin{equation}
\label{Htime}
H_c (\textbf{p}, \textbf{x}, u)= p_0 + p_1 x_2 - p_2 \omega_0^2 u(t),
\end{equation}
translating canonical equations~\eqref{canonicaleqp} into a set of costate equations
\begin{eqnarray}
\label{costate time-1}
\dot{p}_1 &=& 0,
\\
\label{costate time-2}
\dot{p}_2 &=& -p_1.
\end{eqnarray}
Once we solve the costate functions mentioned above, the time-optimal control function $u(t)$ of bang-bang type is obtained as 
\begin{eqnarray}
\label{control function-time bang-ang}
u(t) = \left\{\begin{array}{ll}
0,& t \leq 0
\\
-\delta, & 0 <t <t_1
\\
\delta, & t_1 < t <t_f
\\
0, & t_f \leq t
\end{array}\right.,
\end{eqnarray}
where the minimal time is found to be
\beq
\label{mint}
t^{\min}_f = \frac{2}{\omega_0}\sqrt{\frac{d}{\delta}},
\eeq
with switching point $t_1 =t_f/2$. Fig.~\ref{fig2}(a) illustrates the bang-bang controller $u(t)$, where the parameters are chosen to correspond to the transport experiment of cold atoms~\cite{davidepl2008}, with trap frequency $\omega_0=2\pi\times20$ Hz, transport distance $d=1 \times 10^{-2}$ m, and the mass of  $^{87} \mbox{Rb}$ atoms $m=1.44269 \times 10^{-25}$ kg. Here the constraint on relative displacement $\delta/d=0.1$ is fixed, therefore the minimal time $t^{\min}_f =50.3$ ms. 
However, there exist three sudden jumps in the control function $u(t)$, leading to infinite relative speed of the trap at switching points, which could be problematic in the experimental implementation. 

\subsection{smooth bang-bang control with constrained relative velocity and acceleration}

Motivated by the problem arising from bang-bang control, we introduce more constraints to cancel the sudden jumps, ensuring feasibility in experiments as well. A new component $x_3$ is added into the state vector $\textbf{x}$, with the relations between two nearby components being
\begin{equation}
x_1=q_c,\ x_2=\dot{q}_c,\ x_3=-\omega_0^2 \dot{u}(t),\ u(t)=q_c-q_0.
\label{newnotation}
\end{equation}
Thus, Eq.~\eqref{aeq} can be rewritten into the form for solving the time-optimal control problem as
\begin{eqnarray}
\dot{x}_1&=&x_2,
\\
\dot{x}_2&=&x_3,
\\
\dot{x}_3&=&-\omega_0^2 \dot{u}(t).
\label{aeqnn}
\end{eqnarray}
The new control Hamiltonian $H_c$ (\ref{Hc}) can be updated with the cost functional $J_T$ in Eq.~\eqref{J},
\begin{equation}
\label{Htime}
H_c (\textbf{p}, \textbf{x}, u, \dot{u})= p_0 + p_1 x_2 + p_2 x_3 -  p_3 \omega_0^2 \dot{u}(t),
\end{equation}
giving new costate equations as
\begin{eqnarray}
\label{costate time-1}
\dot{p}_1 &=& 0,
\\
\label{costate time-2}
\dot{p}_2 &=& -p_1,
\\
\label{costate time-3}
\dot{p}_3 &=& -p_2.
\end{eqnarray}
which can be solved easily as $p_1 = c_1$ , $p_2 = -c_1 t +c_2$, and $p_3 = -c_1 t^2/2+c_2 t + c_3$ with constants $c_1$, $c_2$, and $c_3$. Based on Pontryagin's maximum principle~\cite{book}, the time-optimal controller $u(t)$ maximizes the control Hamiltonian (\ref{Htime}) with the new constraint on the relative velocity, $|\dot{u}(t)|\leq\epsilon$.  In order to smooth out the bang-bang control, $\dot{u}(t)$ can be taken as
\begin{eqnarray}
\label{control functiond-time}
\dot{u} (t) = \left\{\begin{array}{ll}
-\epsilon, & 0\ \leq t<t_1
\\
0, & t_1<t <t_2
\\
\epsilon, & t_2< t <t_3
\\
0, & t_3 <t <t_4
\\
-\epsilon, & t_4<t\ {\leq}\ t_f
\end{array}\right..
\end{eqnarray}
After combining the previous constraint on the relative displacement, $|u(t)|\leq\delta$, the ``sudden-jump-free" controller $u(t)$ becomes
\begin{eqnarray}
\label{control function-time}
u(t) = \left\{\begin{array}{ll}
-\epsilon t + c_1, & 0\ \ {\leq}\ t<t_1
\\
c_2, & t_1<t <t_2
\\
\epsilon t + c_3, & t_2< t <t_3
\\
c_4, & t_3 <t <t_4
\\
-\epsilon t + c_5, & t_4<t\ {\leq}\ t_f
\end{array}\right.,
\end{eqnarray}
where $c_2=-c_4=-\delta$, $c_1=0$, $c_3=-(\delta+\epsilon t_2)$, and $c_5=\epsilon t_f$.

\begin{figure}[]
	\includegraphics[width=8.0cm]{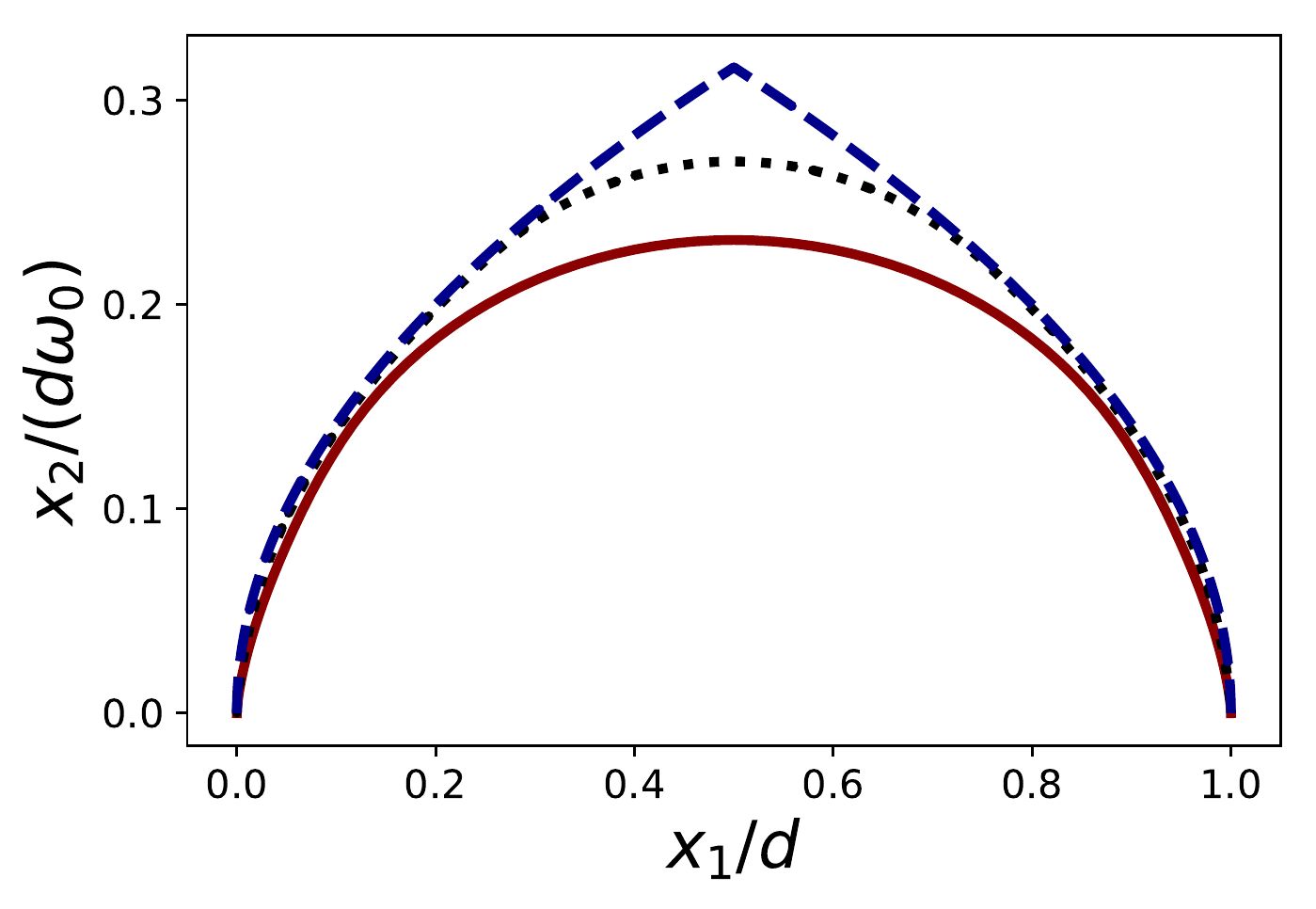}
	\caption{Phase diagram of smooth bang-bang control of fast transport with different relative velocity constraints, where the relative displacement is bounded by $\delta/d=0.1$, keeping the constraint on the relative velocity variable.	The trajectories with $\epsilon/(d\omega_0)=0.05$ (solid red) and  $\epsilon/(d\omega_0)=0.1$ (dotted black) become much smoother with larger time $t_f=68.7$ and  $58.9$ ms, calculated by Eq.~\eqref{eqtfv}, respectively, as compared to the case of bang-bang control (dashed blue) $t_f=50.3$ ms in Fig.~\ref{fig2}. Other parameters are the same as those in Fig.~\ref{fig2}. }  
	\label{fig3}
\end{figure}
According to the boundary conditions, the symmetry, and continuity conditions, one can find four switching points $t_1$, $t_2$, $t_3$ and $t_4$ with the values of $\delta/\epsilon$, $t_f/2-\delta/\epsilon$, $t_f/2+\delta/\epsilon$, and $t_f-\delta/\epsilon$, respectively.
Substituting $u(t)$ into Eq.~\eqref{aeq}, and with boundary conditions [see Eqs.~\eqref{bc1} and (\ref{bc2})], we find
the solution of $q_c(t)$ in different time intervals as follows
\begin{equation}
q_c(t) = \left\{\begin{array}{ll}
\frac{1}{6}\omega_0^2\epsilon t^3
\\
\frac{1}{2}\omega_0^2\delta (t^2 - \frac{\delta}{\epsilon}t + \frac{1}{3}\frac{\delta^2}{\epsilon^2})
\\
-\frac{1}{6}\omega_0^2\epsilon(t-\frac{t_f}{2})^3+\omega_0^2\delta(\frac{t_f}{2}-\frac{\delta}{\epsilon})t-\frac{1}{4}\omega_0^2\delta t_f(\frac{t_f}{2}-\frac{\delta}{\epsilon})
\\
-\frac{1}{2}\omega_0^2\delta[t^2 - (\frac{t_f}{2} - \frac{\delta}{\epsilon})t-\frac{1}{3}\frac{\delta^2}{\epsilon^2}+\frac{t_f^2}{2}]
\\
d- \frac{1}{6}\omega_0^2\epsilon (t_f-t)^3
\end{array}\right.,
\end{equation}
from which the trajectory of the trap center $q_0(t)$ 
can be easily obtained through Eq.~\eqref{aeq}.  
After straightforward calculation, we obtain the near-minimal time as follows,
\begin{equation}
\label{eqtfv}
t_f =\frac{\delta}{\epsilon}
+\frac{2}{\omega_0}\sqrt{\frac{d}{\delta}+\frac{\omega_0^2\delta^2}{4\epsilon^2}},
\end{equation}
which tends to the minimal time in Eq.~\eqref{mint}, when relative velocity is no longer limited, i.e., $\epsilon \rightarrow \infty$. Fig.~\ref{fig2} demonstrates the trajectories of trap center and mass center of the atom with a smoother controller $u(t)$ at switching points, when the relative velocity is bounded.  Apparently, the constraint in relative velocity prolongs the time-optimal transport, as shown in a phase diagram (see Fig.~\ref{fig3}),
where the trajectory becomes smoother. To be precise, the transport time increases from $t_f= 58.9$ to $68.7$ ms,
when the constraint on the relative velocity decreases from $\epsilon/(d \omega_0) =0.1$ to $0.05$,
with the same bounded relative displacement, $\delta/d=0.1$. 
\begin{figure}[]
\includegraphics[width=8.0cm]{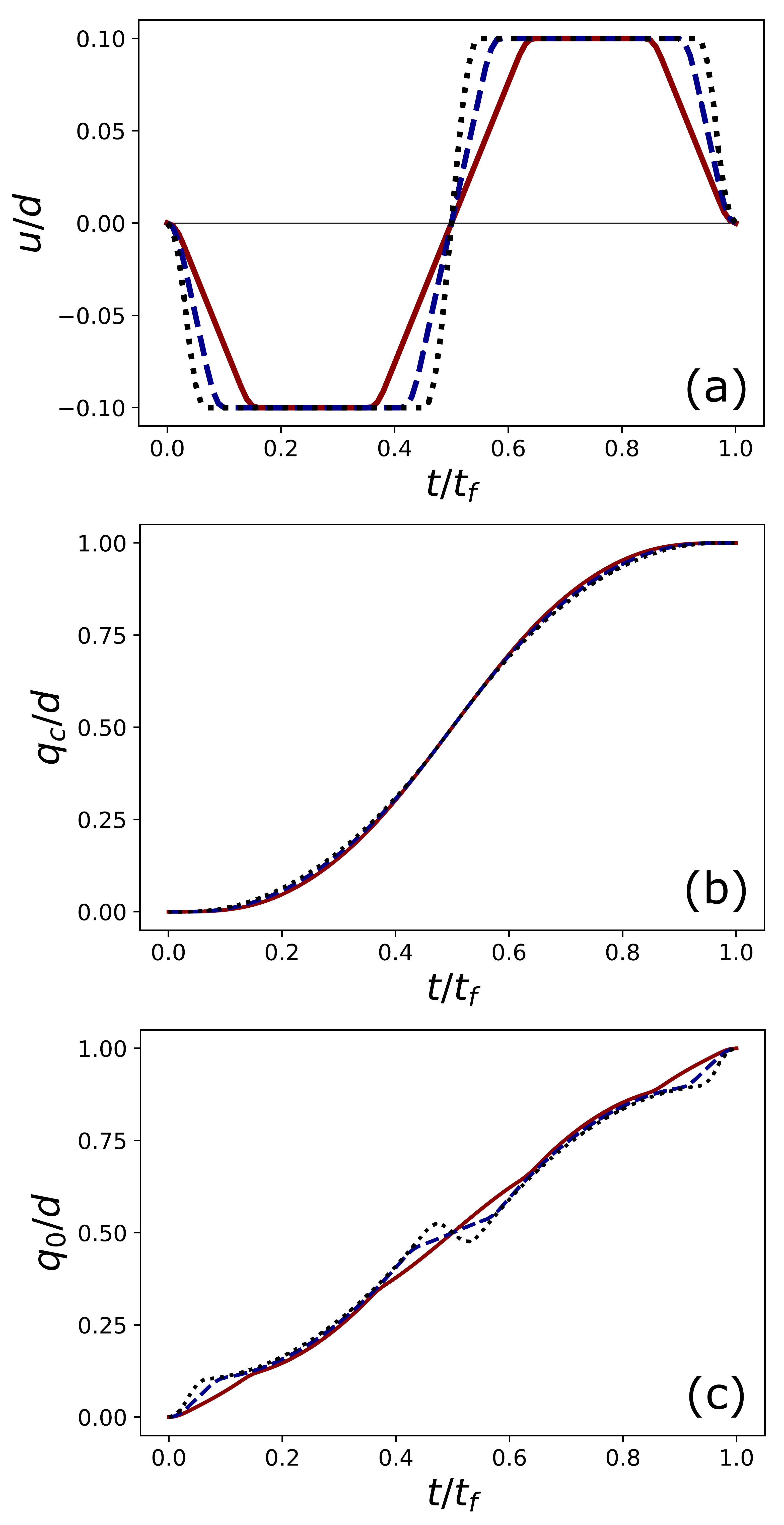}
\caption{(a) Controller $u(t)$ with different constraints, where $\epsilon/(d\omega_0)=0.1$, $\zeta/(d\omega_0^2)=0.5$ (solid red), $\epsilon/(d\omega)=0.2$, $\zeta/(d\omega_0^2)=1$ (dashed blue), $\epsilon/(d\omega_0)=0.5$, $\zeta/(d\omega_0^2)=2$ (dotted black), and other parameters are the same as those in Fig.~\ref{fig2}. In those cases, the near-minimal times are $t_f= 60.5$, $56.1$, and $53.9$ ms, respectively, given by Eq.~\eqref{eqtfa}.  (b,c) The corresponding trajectories of mass center $q_c(t)$ and trap center $q_0(t)$ under different constraints.}
\label{fig4}
\end{figure}

Next, we find the near-minimal-time protocol with an extra constraint condition on the relative acceleration, i.e., $|\ddot{u}(t)|\leq \zeta$, since the discontinuity of trap speed, leads to infinite acceleration in previous protocols. Therefore, the new notation $x_4=\dot{x}_3=-\omega_0^2 \ddot{u}(t)$ is added to equations for defining the control Hamiltonian as 
\begin{eqnarray}
\label{newHtime}
H_c (\textbf{p}, \textbf{x}, u, \dot{u}, \ddot{u})= p_0 + p_1 x_2 + p_2 x_3 + p_3 x_4 - p_4 \omega_0^2 \ddot{u},
\end{eqnarray}
from which we use canonical equation (\ref{canonicaleqp}) to obtain the following costate functions:
\begin{eqnarray}
\label{newcostate time-1}
\dot{p}_1 &=& 0,
\\
\label{newcostate time-2}
\dot{p}_2 &=& -p_1,
\\
\label{newcostate time-3}
\dot{p}_3 &=& -p_2,
\\
\label{newcostate time-4}
\dot{p}_4 &=& -p_3.
\end{eqnarray}
Accordingly, the optimal control that maximizes $H_c$ in Eq.~\eqref{newHtime} is determined by the sign of $p_4$, when $\ddot{u}(t)$ is bounded by $|\ddot{u}(t)| \leq \zeta$. Here we apply three constraints simultaneously, with the other two being $|u(t)| \leq \delta$ and $|\dot{u}(t)| \leq \epsilon$. The near-minimal-time protocol meets the limitations of the relative acceleration, velocity, and displacement simultaneously. As a consequence, the second derivative of the controller, $\ddot{u}(t)$, has the following form of bang-bang type,
\begin{eqnarray}
\ddot{u}(t) = \left\{\begin{array}{ll}
-\zeta, & 0\ \ \  {\leq}\ t<t_1
\\
0, & t_1\ <t <t_2
\\
\zeta, & t_2\ < t <t_3
\\
0, & t_3\  <t <t_4
\\
\zeta, & t_4\ <t<t_5
\\
0, & t_5\  <t <t_6
\\
-\zeta, & t_6\  <t <t_7
\\
0, & t_7\  <t <t_8
\\
-\zeta, & t_8\  <t <t_9
\\
0, & t_9\  <t <t_{10}
\\
\zeta, & t_{10} <t \leq t_f
\end{array}\right.,
\end{eqnarray}
With boundary conditions at switching points, after a simple integration, $\dot{u}(t)$ can be given by
\begin{eqnarray}
\dot{u}(t) = \left\{\begin{array}{ll}
-\zeta t, & 0\ \ \ {\leq}\ t<t_1
\\
-\epsilon, & t_1\ <t <t_2
\\
\zeta(t-t_2)-\epsilon, & t_2\ < t <t_3
\\
0, & t_3 \ <t <t_4
\\
\zeta(t-t_4), & t_4\ <t<t_5
\\
\epsilon, & t_5 \ <t <t_6
\\
-\zeta(t-t_6)+\epsilon, & t_6 \ <t <t_7
\\
0, & t_7 \ <t <t_8
\\
-\zeta(t-t_8), & t_8 \ <t <t_9
\\
-\epsilon, & t_9 \ <t <t_{10}
\\
\zeta(t-t_{10})-\epsilon, & t_{10} <t \leq t_f
\end{array}
\right.,
\end{eqnarray}
from which the switching points can be calculated as
$t_1=\epsilon/\zeta$,
$t_2=\delta/\epsilon$,
$t_3=\delta/\epsilon+\epsilon/\zeta$,
$t_{4,5}=\frac{1}{2}(t_f-2\delta/\epsilon \mp \epsilon/\zeta)$,
$t_{6,7}=\frac{1}{2}(t_f+2\delta/\epsilon \mp \epsilon/\zeta)$,
$t_8=t_f-t_3$,
$t_9=t_f-t_2$,
and $t_{10}=t_f-t_1$.
With these switching points, the controller [see Fig.~\ref{fig4}(a)] can be finally expressed by
\begin{eqnarray}
\label{newcontrol function-time}
u(t) = \left\{\begin{array}{ll}
-\frac{1}{2}\zeta t^2 
\\
-\epsilon(t-\frac{\epsilon}{\zeta})-\frac{\epsilon^2}{2\zeta}
\\
\frac{1}{2}(\zeta t^2-2\epsilon t+\frac{\epsilon^2}{\zeta}-\frac{2\delta\zeta t}{\epsilon}+\frac{\delta^2\zeta}{\epsilon^2})
\\
-\delta
\\
-\delta+\frac{[\epsilon^2+2\delta\zeta-\epsilon(t_f-2t)\zeta]^2}{8\epsilon^2\zeta}
\\
\epsilon(t-\frac{t_f}{2})
\\
\frac{1}{8}\left[-\frac{4\delta^2\zeta}{\epsilon^2}-\frac{(\epsilon+\zeta(t_f-2t))^2}{\zeta}+\frac{4\delta(\epsilon-\zeta t_f+2\zeta t)}{\epsilon}\right]
\\
\delta
\\
\delta-\frac{(\epsilon^2+\delta\zeta+\epsilon\zeta(t-t_f))^2}{2\epsilon^2\zeta}
-\frac{\epsilon[\epsilon+2(t-t_f)\zeta]}{2\zeta}
\\
\frac{1}{2}\zeta(t_f-t)^2
\end{array}\right.,
\end{eqnarray}
\begin{figure}[]
	\includegraphics[width=8.0cm]{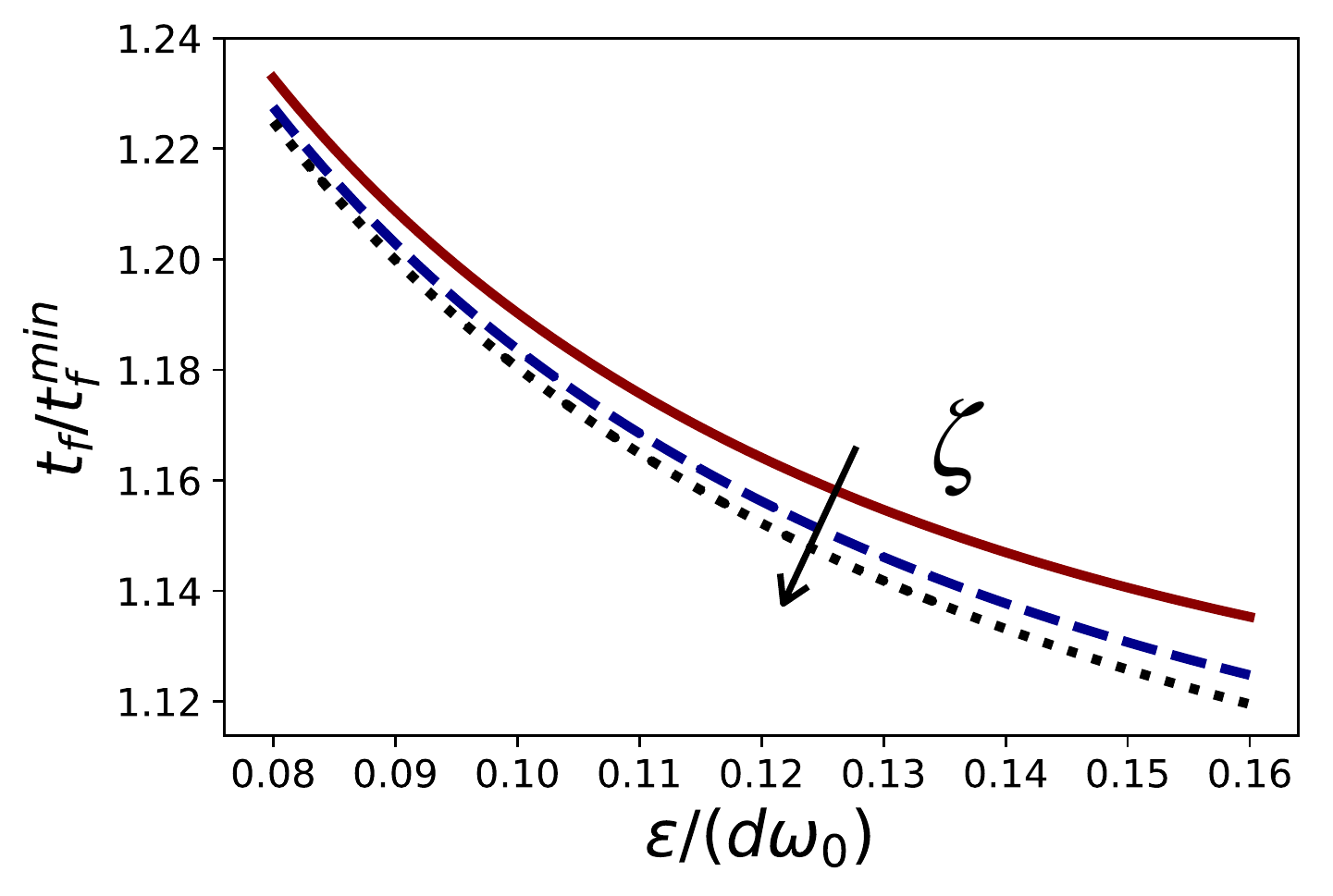}
	\caption{ Dependence of near-minimal time $t_f$ on differently bounded relative velocities and accelerations: $\zeta/(d\omega_0^2)=0.8$ (solid red), $\zeta/(d\omega_0^2)=1.2$ (dashed blue), and  $\zeta/(d\omega_0^2)=1.6$ (dotted black), where other parameters are the same as those in Fig.~\ref{fig2}.}
	\label{fig5}
\end{figure}
Trajectories of trap center $q_0(t)$ and mass center of a cold atom $q_c(t)$ can be easily calculated through Eq.~\eqref{aeq} [see Fig.~\ref{fig4}(b) and (c)]. 
Obviously, this shows a feasible way to realize smooth transport, only taking a little more time as cost than the previous cases. Thus, the final expression of near-minimal time in this case is given by
\begin{equation}
\label{eqtfa}
t_f =\frac{\delta}{\epsilon} + \frac{\delta \epsilon}{\zeta}
+\frac{2}{\omega_0}\sqrt{\frac{d}{\delta}+\frac{\omega_0^2\delta^2}{4\epsilon^2}}.
\end{equation}
The minimal time given here is just increased a little by  $\delta\epsilon/\zeta$, which is the exact price for smooth bang-bang control by bounding relative acceleration. For instance, we choose three different 
constraints in Fig.~\ref{fig4}, where $\epsilon/(d\omega_0)=0.1$,
$\zeta/(d\omega_0^2)=0.5$ (solid red),
$\epsilon/(d\omega)=0.2$, $\zeta/(d\omega_0^2)=1$ (dashed blue), $\epsilon/(d\omega_0)=0.5$, $\zeta/(d\omega_0^2)=2$ (dotted black), and other parameters are the same as those in Fig.~\ref{fig2}. It is obvious that the larger the constraint, the more similar the control. Experimental realization without energy excitation also becomes harder for larger constraints despite near-minimal times; here we emphasize that one can further smooth the protocol by introducing more constraints on the higher-order derivatives of the controller. However, it might be unnecessary to do so, since numerical studies given below convince us.

Figure \ref{fig5} clarifies how much price one should pay for smoothing the bang-bang control out. In general, the influence of the constraint on the relative velocity, $\epsilon$, is more pronounced, as compared to the constraint on the relative acceleration, $\zeta$. Setting more constraints on its first and second derivatives of the controller can smooth out bang-bang time-optimal control more, with extra cost of transport time as a trade-off. 

\begin{figure}[]
\includegraphics[width=8.0cm]{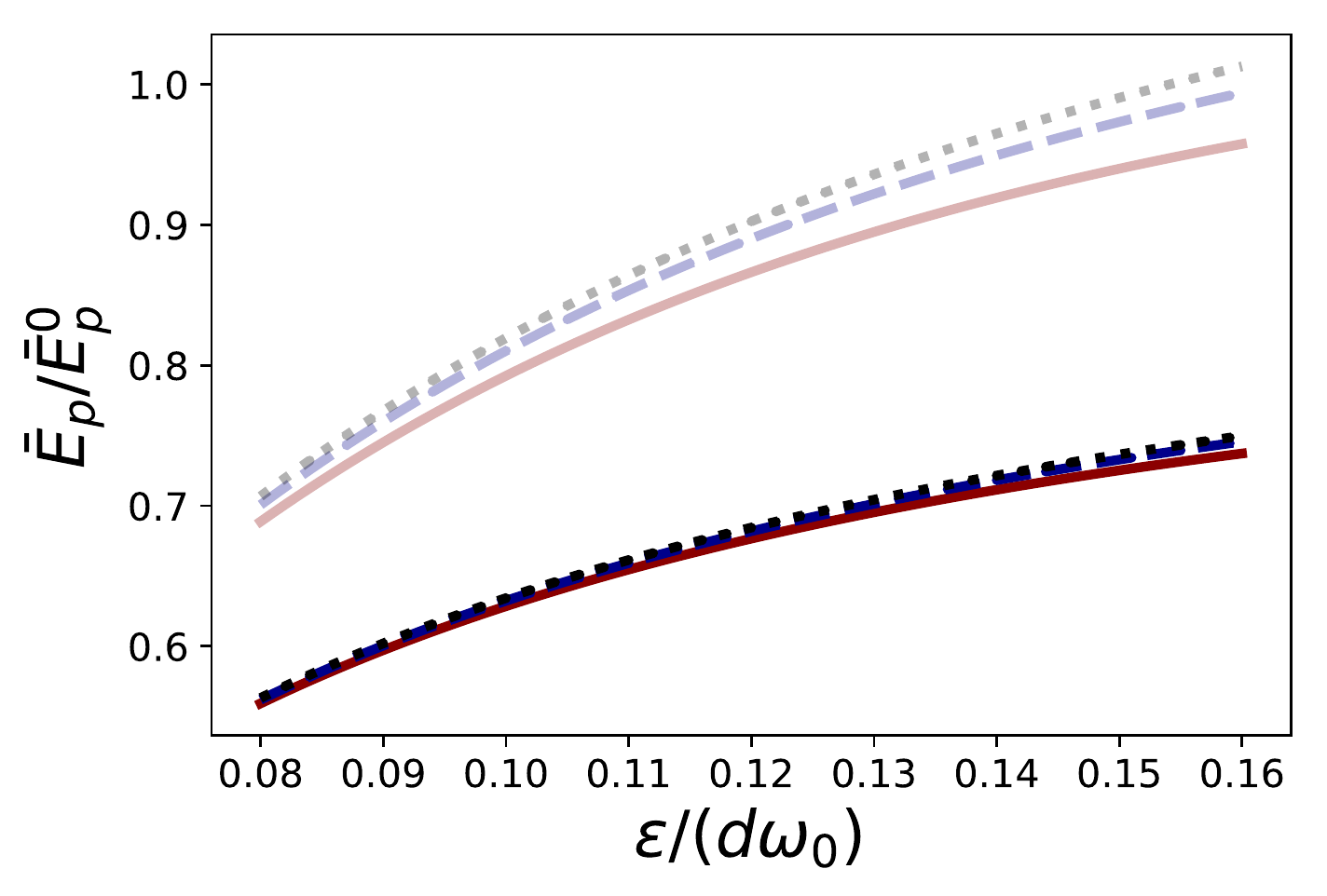}
\caption{Time-averaged potential energy $\bar{E}_p$ which characterizes energy excitation of the transport, rescaled by $\bar{E}_p^0$ (time-optimal bang-bang control). We calculate $t_f$ for Eq.~\eqref{poly} with the same constraints for a fair comparison, resulting in different controllers and time-averaged potential energies. Our smooth bang-bang protocols excite much less than the higher-order polynomial \textit{Ans\"atze}~\eqref{poly} (faded lines). Near-minimal time $t_f$ and other parameters are the same as those in Fig.~\ref{fig5}.}
\label{fig6}
\end{figure}

Moreover, we shed light on the energy excitation, characterized by time-averaged potential energy (\ref{energy1}), for smooth bang-bang protocols and a polynomial trajectory ~(\ref{poly}) that is used in the experiment \cite{nessnjp2018}. In general, the energy excitation  can be suppressed by smooth bang-bang protocols (see Fig.~\ref{fig6}) since it is proportional to $u^2$. In spite of the fact that the excitation energy increases with the upper bounds of velocity and acceleration, a fair comparison would be to calculate the transport time of polynomial trajectory (\ref{poly}) corresponding to each smooth bang-bang protocol with different upper limits. Clearly, the polynomial trajectory produces larger energy excitation than smooth bang-bang protocol.  Also, one can calculate the sloshing amplitudes, to quantify the performance of STA. The ultimate sloshing can be suppressed from $\mathcal{A} (t_f) =4 \times 10^{-3}$ to $\mathcal{A}(t_f)\simeq10^{-12}$ by smooth bang-bang controls. As mentioned above,  the polynomial \textit{Ans\"atz} \eqref{poly} carried out in the experiment \cite{nessnjp2018} is not optimized enough with respect to time or time-averaged potential energy, though the corresponding sloshing is $\mathcal{A}(t_f)\simeq10^{-16}$.  In this case, the resulting relative displacement during the process is continuous, but exceeds the upper limit, $|u|\leq\delta$, used in the time-optimal solution. This might be problematic in practice when the anharmonic effect is taken into account in an optical Gaussian trap \cite{qipra2015,qijpb2016,jing}.

\section{Numerical multiple shooting algorithm}

In this section, we present the numerical multiple shooting method to solve such near-time-optimal control with two-fold reasons. On one hand, the analytical expressions become too complicated to solve, when high-order derivatives of controller are considered. Thus, the numerical algorithm is required to calculate automatically the switching points and minimal time with different constraints for double checking and simplicity. On the other hand, a shooting method may encounter numerical difficulties for solving the optimal control, since the shooting function is not smooth when the control is bang-bang \cite{bassam}. Beyond that, the reason for applying multiple shooting method, as a tool of our numerical studies, is that, it can be parallelized for certain problems, which can have a non-negligible advantage in efficiency, comparing with other algorithms. In what follows, we shall formulate the boundary-value problem, and solve the smoothing procedure by using multiple shooting method. The detailed steps of our algorithm are as follows.

\begin{figure}[]
\includegraphics[width=8.0cm]{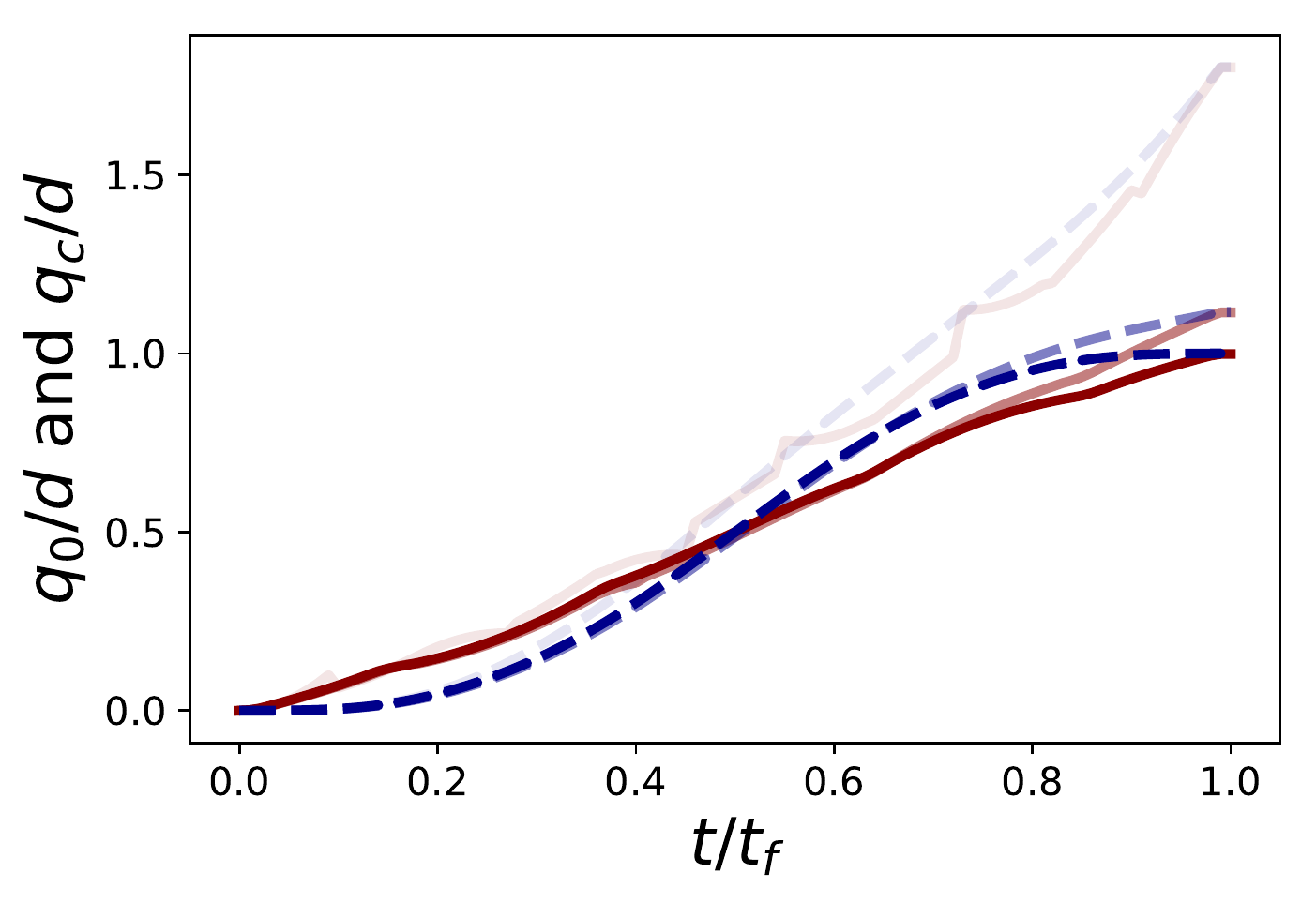}
\caption{With updating rate $\rho=0.5$, the incorrect initial guess (upper faded lines) of switching points $\{5, 10,...,55\}$ (ms) converges to a smooth bang-bang solution of near-time-optimal transport quickly (middle faded lines, after three epochs), with the constraint conditions $\delta/d=0.1$, $\epsilon/(d\omega_0)=0.1$, and $\zeta/(d\omega_0^2)=0.5$, by iteration number of 13. The final correct trajectories of mass center $q_c(t)$ and trap center $q_0(t)$ are presented by red solid and dashed blue lines, respectively. Other parameters are the same as those in Fig. \ref{fig2}.}
\label{fig7}
\end{figure}

(\romannumeral1) We get the expression of $q_c(t)$ with ten switching points and the minimal transport time, which are unknown, by solving the classical equation with boundary conditions and continuous conditions.

(\romannumeral2) Then we can write a column vector $f=(q_c(t_f)-d,\dot{q_c}(t_f),u(t_f),\dot{u}(t_f),u(t_3)+\delta,u(t_7)-\delta,\dot{u}(t_1)+\epsilon,\dot{u}(t_3),\dot{u}(t_5)-\epsilon,\dot{u}(t_7),\dot{u}(t_9)+\epsilon)^{T}$. Its norm, as the objective function, should be optimized to zero when all the switching points and minimal time are corrected.

(\romannumeral3) A Jacobian matrix $J_{ij}=\partial f_i/\partial t_j$ is defined to calculate the modifications of switching points and minimal time.

(\romannumeral4) We set another column vector $g$ to be $g=(t_1,t_2,t_3,t_4,t_5,t_6,t_7,t_8,t_9,t_{10},t_f)^T$. All the elements' initial values are our assumptions of switching points and minimal transport time, which will be updated by the algorithm iteratively.

(\romannumeral5) Calculate the values of $f$ and $J$ with times given by $g$. Gradient Del is defined as $\text{Del}=J^{-1}f$. In this way, the modified $g$ will be $g=g-\rho \text{Del}$, where $\rho$ is a constant deciding the speed of convergence between zero and one. After that, calculate the norm of the new $f$.

(\romannumeral6) Repeat step (\romannumeral5) until the norm of $f$ is smaller than an acceptable fixed tolerance value.

To demonstrate the algorithm, we give an example of smooth transport calculated with multiple shooting method and plot the incorrect initial guess, middle (epoch=3), and final trajectories (epoch=13), for showing how the protocol converges to the near-time-optimal solution (see Fig.~\ref{fig7}), where the constraints on the relative displacement, velocity, and acceleration---$\delta/d=0.1$, $\epsilon/(d\omega_0)=0.1$, and $\zeta/(d\omega_0^2)=0.5$, respectively---hold.

\begin{figure}[]
\includegraphics[width=8.0cm]{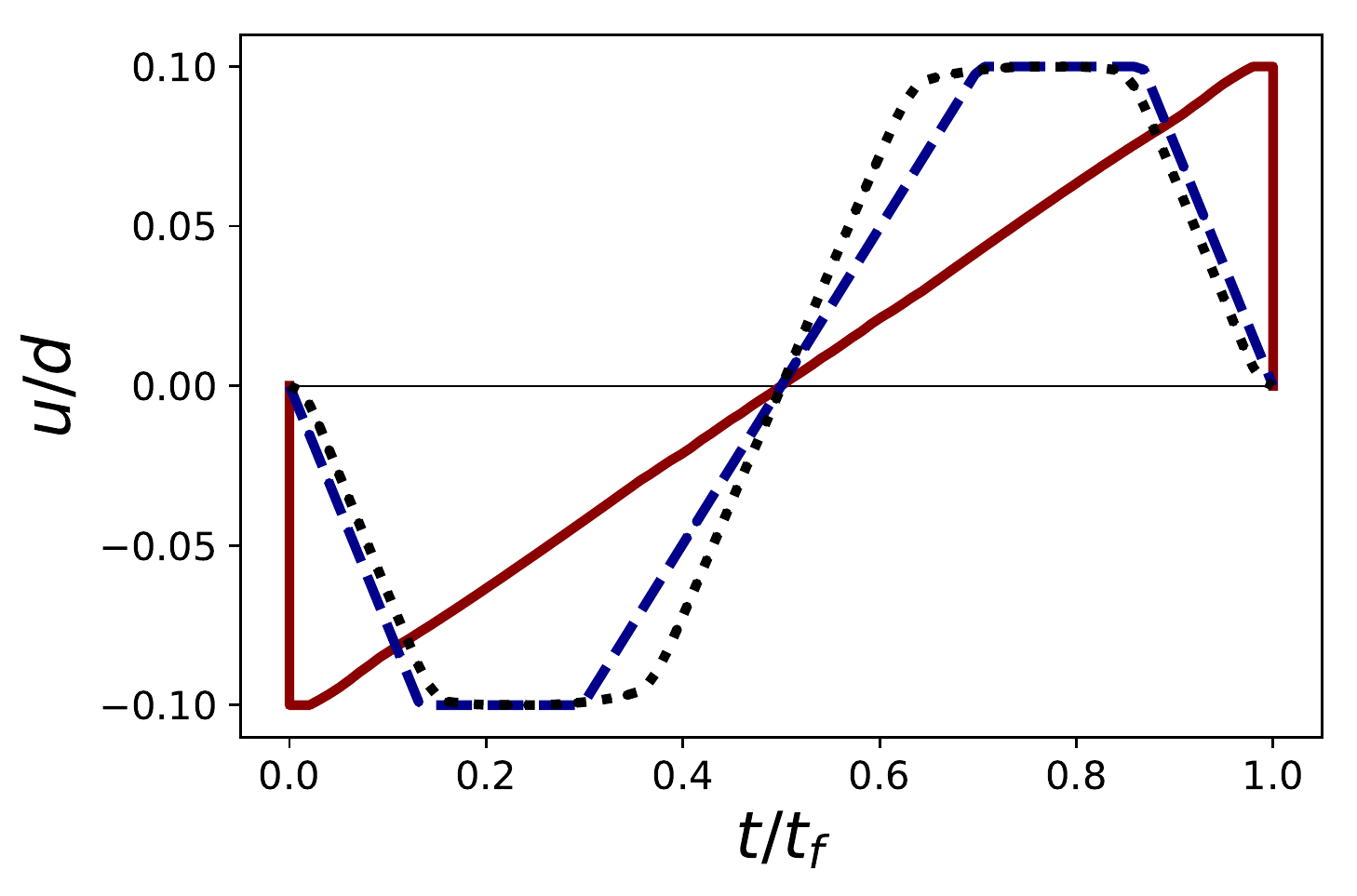}
\caption{Controller $u(t)$ that minimizes the time-averaged potential energy $\bar{E}_p$ within a fixed  time  $t_f=60$ ms. Controller $u(t)$ is bounded by $\delta/d=0.1$ (solid red); $\delta/d=0.1$ and $\epsilon/(d\omega_0)=0.1$ (dashed blue); and $\delta/d=0.1$, $\epsilon/(d\omega_0)=0.1$ and $\zeta/(d\omega_0^2)=0.5$ (dotted black). The corresponding time-averaged potential energies $\bar{E}_p/\bar{E}^{\min}_p$ are $1.0002$, $1.4918$, and $1.6099$, which are larger than the lowest bound for the potential energy in unbounded control. $N=100$, $M=10$, and other parameters are the same as those in Fig.~\ref{fig2}.}
\label{fig8}
\end{figure}

In addition to time-optimal control, the minimization of energy excitation is dealt with using the same numerical algorithm mentioned above. From Eq.~(\ref{energy1}), the cost functional reads
\begin{equation}
J_E=\int_0^{t_f} E_pdt=\int_0^{t_f}\frac{1}{2}m\omega_0^2u^2dt,
\end{equation}
where transport time $t_f$ is fixed. The control Hamiltonian can be written as
\begin{equation}
H_c=-p_0\frac{1}{2}m\omega_0^2u^2+p_1x_2-p_2\omega_0^2u,
\end{equation}
leading to new costate equations. Following Pontryagin's maximum principle, the unbounded control, i.e., without any constraints on $u$, gives the lowest bound~\cite{chenpra2011}
\beq
\bar{E}^{\min}_p = 6md^2/\omega^2_0 t^4_f,
\eeq 
with linear time-varying controller 
\beq
u(t) =\frac{6d}{\omega^2_0 t^2_f} \left(2\frac{t}{t_f}-1\right).
\eeq
Similarly, the controller $u(t)$ is not zero at $t=0$ and $t=t_f$, implying an infinite speed of the moving trap. The more complicated case with bounded controller $u(t)$ can be calculated as well in Ref.~\cite{chenpra2011}. However, it is impossible to achieve the analytical expression,  when higher-order derivatives of controller $u(t)$ are bounded. For this task, we apply multiple shooting method again to numerically design STA with arbitrary dimension of states and constraints. Transport interval $[0,t_f]$ is partitioned by $N$ grid points, where the control function consists of $N-1$ subintervals with length $t_f/(N-1)$. In order to find the solution of boundary-value problems, we define a $D$-dimensional state $\textbf{x}$ and its derivative $\dot{\textbf{x}}$, initializing it by guessing. We define a time step $h=t_f/(N-1)(M-1)$ for applying fourth-order Runge-Kutta method as an ordinary differential equation (ODE) solver. In each subinterval, we calculate the following four terms
\begin{eqnarray}
\textbf{k}_1&=&\dot{\textbf{x}}_j,\\
\textbf{k}_2&=&\dot{\textbf{x}}_j+\frac{h}{2}\dot{\textbf{k}}_1,\\
\textbf{k}_3&=&\dot{\textbf{x}}_j+\frac{h}{2}\dot{\textbf{k}}_2,\\
\textbf{k}_4&=&\dot{\textbf{x}}_j+h\dot{\textbf{k}}_3,
\end{eqnarray}
for updating the state of the next time step by
\begin{equation}
\textbf{x}_{j+1}=\textbf{x}_{j}+\frac{h}{6}(\textbf{k}_1+2\textbf{k}_2+2\textbf{k}_3+\textbf{k}_4),
\end{equation}
where $j\in\{1,2,...,M-1\}$. Thus, we obtain $\textbf{x}_{i+1}$ for all $i\in\{1,2,...,N-1\}$ with a given $\textbf{x}_{i}$. Combining it with an optimizer, we can optimize any objective function, satisfying constraint conditions at the same time. In Fig.~\ref{fig8}, we use multiple shooting method for solving ODEs, minimizing potential energies with \textsc{matlab} optimizer \textsc{fmincon} under different constraint conditions. Again, the linear time-varying controller $u(t)$, initially with drastic changes at initial and final times, becomes smoother at the cost of potential-energy increase. 

\begin{figure}[]
	\includegraphics[width=8.0cm]{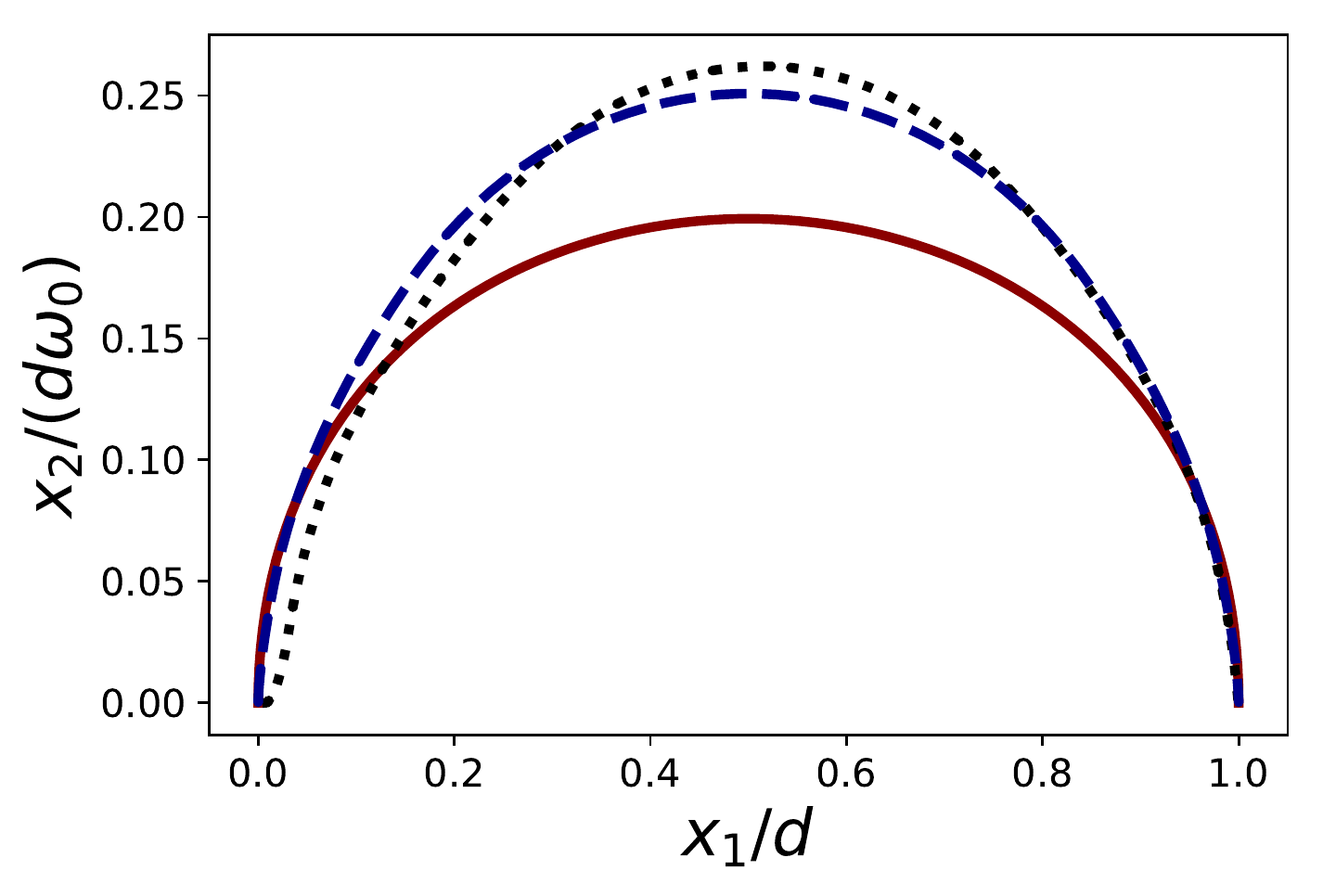}
	\caption{Phase diagram of smooth bang-bang control of fast transport within fixed $t_f$, minimizing the time-averaged potential energy $\bar{E}_p$ numerically. Constraint conditions and other parameters are the same as those in Fig.~\ref{fig8}.}
	\label{fig9}
\end{figure}

Moreover, we show the phase diagram of smooth transport protocols in Fig.~\ref{fig9}. We notice that when higher-order constraints are introduced the phase diagram becomes asymmetric around $t=t_f/2$, resulting in a local minimum of potential energy. It is hard to obtain a global optimal solution because the gradient algorithm depends on its initial input as trial solution. However, with a reasonable range of guesses, this numerical algorithm converges to sub-optimal solutions, which are friendly enough for experimental implementations.

\section{Conclusion and Outlook}
In summary, we present analytical and numerical methods for 
smooth bang-bang shortcuts to adiabaticity for atomic transport. Preceding researches provide the time-optimal solution, as typical bang-bang control, which contains drastic changes of controllers, resulting in high residual energy and difficulty in experimental implementation. Here we propose smooth bang-bang controls, corresponding to the near-minimal time, by bounding the first and second-order derivatives of the controller. Further comparison between our smooth bang-bang and simple polynomial protocols shows that both energy excitation and sloshing amplitude are significantly suppressed with increasing slightly the transport time as a tradeoff. To make our results more applicable, the numerical multiple shooting algorithm is developed for time or energy minimization, where an analytical solution might not be feasible or solvable. Within this framework, different \textit{Ans\"atze}, including high-order polynomial and trigonometric functions, can be compared, and they are suitable for obtaining sub-optimal solutions~\cite{compare} and enhanced STA~\cite{qcesta} in further work.

Finally, we emphasize that our analytical and numerical methods, supplemented by machine learning~\cite{shersonnature2016,sels,machinelearning}, will provide a versatile toolbox for quantum control since time-optimal bang-bang solutions are ubiquitous with applications including atom cooling~\cite{stefanatospra2010,kosloff,xiaojingpra2014}, transport of trapped-ion qubits \cite{mikelpra2013,mikelpra2014,xiaojing14,xiaojing15,xiaojing18}, ground-state preparation~\cite{freericksbangbang,freericksbangbang2}, and long-distance transport in an optical lattice~\cite{alberti,xiaojing,andreas}. These results can be further extended to other problems, including compact interferometers with spin-dependent force~\cite{socBEC,armsguiding},
load manipulation by cranes in a classical system~\cite{mugacrane}, and Brownian motion in statistical physics~\cite{prados}.

\section*{Acknowledgement}

This work is partially supported by National Natural Science Foundation of China (Grant No.~11474193), STCSM (Grants No.~2019SHZDZX01-ZX04, No.~18010500400, and No.~18ZR1415500), the Program for Eastern Scholar, and HiQ funding for developing shortcuts to adiabaticity (Grant No.~YBN2019115204). X.C. also acknowledges Ram\'on y Cajal program of the Spanish MCIU (Grant No.~RYC-2017-22482), QMiCS (Grant No.~820505), and OpenSuperQ (Grant No.~820363) of the EU Flagship on Quantum Technologies, Spanish Government Grant No.~PGC2018-095113-B-I00 (MCIU/AEI/FEDER, UE), Basque Government Grant IT986-16, as well as EU FET Open Grant Quromorphic.

\end{document}